# From GWAS to transcriptomics in prospective cancer design - new statistical challenges.

Sandra Plancade[1,*], Gregory Nuel[2] and Eiliv Lund[1,2]

(1) Department of Community Medicine, Faculty of Health Sciences University of Tromsø, 9037 Tromsø, Norway
(2) Department of Applied Mathematics, MAP5, 45 rue des Saints-Pères, University Paris Descartes, 75006 Paris

(*) Corresponding author

email: Sandra Plancade - sandra.c.plancad@uit.no
Gregory Nuel - gregory.nuel@parisdescartes.fr
Eiliv Lund - eiliv.lund@uit.no

**Running title:** Transcriptomics in prospective cancer design.

**Fundings:** Grant ERC-2008-AdG 232997-TICE "Transcriptomics in cancer epidemiology".

**Corresponding Author:**     Sandra Plancade
Institut Curie, 26 rue d'Ulm,
75248 Paris cedex 05
Phone: +33 (0)1 55 43 14 63
Fax: +33 (0)1 55 43 14 69

**Word count:** 3489

**Number of figures:** 4.

**Abstract.** With the increasing interest in post-GWAS research which represents a transition from genome-wide association discovery to analysis of functional mechanisms, attention has been lately focused on the potential of including various biological material in epidemiological studies. In particular, exploration of the carcinogenic process through transcriptional analysis at the epidemiological level opens up new horizons in functional analysis and causal inference, and requires new designs together with adequate analysis procedures. In this article, we present the post-genome design implemented in the NOWAC cohort as an example of a prospective nested case-control study built for transcriptomics use, and discuss analytical strategies to explore the changes occurring in transcriptomics during the carcinogenic process in association with questionnaire information. We emphasize the inadequacy of survival analysis models usually considered in GWAS for post-genome design, and propose instead to parameterize the gene trajectories during the carcinogenic process. This novel approach, in which transcriptomics are considered as potential intermediate biomarkers of cancer and exposures, offers a flexible framework which can include various biological assumptions.

# 1 Introduction

Classical epidemiology investigates the association between environmental variables and a disease of interest at the population level. During the past decades, research has moved to molecular epidemiology which incorporates biological material together with more classical questionnaire data. In particular, the development of microarray technologies has made possible agnostic studies including genome-wide data [1]. Prospective GWAS (Genome Wide Association Studies) analyse the association between genomics data, in particular SNPs (Single Nucleotide Polymorphisms), and the time of appearance of a disease through prediction and risk estimation.

To address the important issue of bias selection of cases and controls in epidemiology, cohort designs were established in the 50's whereby a large number of individuals are recruited at the beginning of the study and followed either individually or through a register. The cases arising during the follow-up can be compared with the controls in the cohort in an unbiased way. When the variables of interest are expensive or complicated to collect, a limited number of controls are randomly selected in the cohort for each case, termed a nested case-control design, and the analysis is based on measurements of the variables for nested cases and selected controls.

The statistical analysis of prospective GWAS is usually based on survival analysis models which model failure time as a function of genomics and lifestyle variables. In particular, the Cox proportional hazard rate model [2], which is the corner stone of many GWAS, was developed in the early 70's in order to estimate relative risk without computing the time baseline through the minimization of a partial likelihood. Different types of penalization have been introduced to compensate for large dimension covariates [3, 4]. Thomas [5] adapted the likelihood to integrate nested case-control data. Furthermore, a huge number of statistical articles propose generalizations of the Cox model, in particular to include time-varying coefficients [6, 7] or covariates measured with errors [8].

With the increasing interest in post-GWAS research which represents a transition from genome-wide association discovery to analysis of functional mechanisms, attention has been lately focused on the potential of including other biological material (mRNA, miRNA, methylation etc) in epidemiological studies to move from prediction and risk estimation to causal inference [9]. To address these novel issues, cohorts including various type of biological material have been established. In particular, the Norwegian Women And Cancer (NOWAC) post-genome cohort [10] which incorporates transcriptomics measurements (mRNA) in peripheral blood together with lifestyle information collected in a prospective design offers the opportunity to explore the functional changes associated with the carcinogenic process on transcriptional outputs .

In this paper, we will discuss analytical strategies to explore the changes occurring in transcriptomics in peripheral blood during the carcinogenic process in association with questionnaire information based on a prospective nested case-control design. This design differs from the more classical genome-wide studies, which can be classified in two categories. Cross-sectional or hospital case-control studies including transcriptomics or genomics data, which do not involve a time aspect, [11] and prospective GWAS including genomics data, in which individuals are followed over time. The post-genome design which incorporates transcriptomics data in a prospective design represents a hybrid between these two categories.

Since the post-genome design shows high similarities with the classical prospective GWAS, epidemiologists could be tempted to apply statistical methods developed for GWAS nested case-control data to this new type of study. Nevertheless, the distinct nature of the association between cancer and genomics in prospective GWAS, and cancer and transcriptomics in the post-genome cohort calls for a different statistical approach. Instead of the survival analysis model usually considered in GWAS:

$$P[T|G,E] \qquad (1)$$

with $T$ the time of failure, $G$ genomics data, and $E$ environmental exposures, we propose to parameterize the trajectory of the gene expression during the carcinogenic process:

$$P[G|T,E] \qquad (2)$$

with *T* the time before diagnosis, *G* transcriptomics data and *E* environmental exposures. We will emphasize that despite the connection between (1) and (2), the statistical approaches based on these two quantities target a different nature of associations between genes and cancer, and that the second approach seems more suitable and more directly interpretable to explore the functional changes occurring in gene expression during the carcinogenic process.

## 2  Transcriptomics in a prospective nested case-control design

We present the post-genome design implemented in the NOWAC post-genome cohort [10] as an example of a prospective nested case-control study built for transcriptomics use. The NOWAC post-genome cohort includes 49,633 women born during 1943-57 who gave a blood sample and filled-in a detailed questionnaire regarding various aspects of their lifestyle at the beginning of follow-up. The blood samples were stored in conditions that preserve the mRNA (PAX-tubes). The women are followed by linkage with the Norwegian Cancer Register, and for each woman in the cohort diagnosed with cancer, a control is randomly picked-up among the healthy women with same age in the cohort (Figure 1). Blood samples from both case and control are analysed on the same Illumina WG6 or HT12 BeadChip. During 7 years of follow-up, 739 incident breast cancer were identified.

### 2.1  Exploration of functional changes in peripheral blood transcriptome

A challenging issue raised by post-genome research is the exploration of functional changes associated with carcinogenesis. In this context, the NOWAC post-genome cohort explores the potentiality of using transcriptomics as biomarkers of the carcinogenic process, by examining how gene expression is affected by carcinogenesis.

Let us consider a hypothetical example to illustrate our approach. For a given case, assume that, the mean expression of a gene *g* has the trajectory presented on the first column of Figure 2. The ideal design to recover the gene trajectory would consist of longitudinal data with multiple time measurements for the same individual. But this type of data, available in clinical studies, would be much more expensive and complicated to collect in a cohort. Therefore, the measurements taken at different times before diagnosis on different cases are incorporated to mimic a longitudinal design. Assume that gene $g$ has the five trajectories presented on the second column of Figure 2 (dotted lines) for five cases in the cohort. Then, single measurements of gene expression for each case at different times before diagnosis enable recovery of the trend of the trajectory (third column of Figure 2).

In the hypothetical example above, the trajectories of gene $g$ are similar for all individuals. But, in epidemiological studies, gene expression is subject to important variations from one individual to the other. Our goal is to detect a general trend due to carcinogenesis in all cases or in a pre-defined subset of cases (individual with a given type of cancer, etc), but high variations between individuals will dilute the effect of carcinogenesis. Several epidemiological studies have previously assessed the effect of environmental exposures on gene expression [12-14] Therefore, the exposure variables extracted from the questionnaire are incorporated in the analysis in order to partially correct individual variations and increase the sensitivity of detection of genes involved in the carcinogenic process.

## 2.2  From GWAS to post-genome design: a different point of view

The point of view adopted in classical prospective GWAS studies is the following: given the values of some covariates (genomics and environmental), what is the risk of developing a cancer at some time? In this context, genomics and exposures variables are considered as risk factors for cancer.

In the post-genome design, we analyse how transcriptomics is affected by both carcinogenesis and exposures. Transcriptomics are assumed to contain some information which could potentially represents intermediate biomarkers of cancer and exposures. This change in point of view is illustrated in Figure 3. The first column presents the classical nested case-control point of view: we observe the time of failure of all individuals given some variables. The second column presents the novel point of view adopted for the post-genome design: the time of reference is not the beginning of the study but the time of diagnosis.

**Role of exposures.** The post-genome design includes two types of environmental or lifestyle exposures extracted from the questionnaire:

(i) Exposures possibly associated with breast cancer (carcinogens): in particular, risk factors related to the history of the individual (age of first pregnancy, use of hormone replacement therapy etc.).

(ii) Exposures not associated with breast cancer but which may affect gene expression: in particular, diet, consumption of tobacco, use of medication, at time of blood sample.

Classical epidemiological studies which focus on prediction and risk estimation only includes the exposures from the first category. Conversely, in post-genome studies, the second type of exposures are considered as sources of noise on transcriptional outputs and should be included in the analysis to improve sensitivity. Thus, the study of gene trajectories in transcriptomics would be optimized by incorporation of exposures that affect gene expression, regardless of their association to cancer.

**Role of controls.** In classical nested case-control GWAS, controls are considered as censored observations at the time of failure of the matched case: if the control is selected at time $t$, the only information used is that the control was not diagnosed with cancer before $t$. Inclusion of controls in the design is essential in order to identify risk factor and estimate relative risks.

The situation is different in post-genome cohort: a priori, the exploration of gene trajectories as a function of time to diagnosis would not require the inclusion of controls. Nevertheless, controls are considered as references in order to remove technical and age effects. Indeed, the pairs are matched by age, therefore considering the difference of log-expression between matched case and control should remove an additive effect of age on gene expression. Similarly, since matched case-control pairs are analysed on the same BeadChip, the case-control difference reduces technical noise.

However, if a control is close to diagnosis at time of matching, the carcinogenesis effect in the case transcriptome could be eliminated by the case-control difference. Thus, our analysis is based on the implicit assumption that the control is not in the "final phase" of carcinogenic process at time of blood sample. In the NOWAC post-genome cohort which explores carcinogenesis 7 years before diagnosis, this assumption amounts to considering that the controls will remain cancer-free 7 years after the blood sample. This approximation is reasonable for two reasons: First of all, the incidence rate along the 7 years of follow-up is small (739/49633 = 0.015) so only a few controls would be implicated. Moreover, the follow-up through the Norwegian Cancer Register enables removal of the pair if the control develops a cancer in the future.

# 3 Statistical approach for transcriptomics in nested-case control prospective design

In the previous section, we underlined the differences between classical prospective GWAS and the post-genome design. The novel challenges raised by the transition from risk estimation and prediction from genomics to exploration of functional changes on transcriptomics naturally lead us to consider longitudinal models (2) with $G$ in alternative to the classical survival analysis models (1) used in GWAS. Nevertheless, considering survival analysis models to detect genes whose expression changes during the carcinogenic process could appear natural: if the expression of a gene depends on time to diagnosis, this gene should be involved in (1). Nevertheless, even if models (1) and (2) are related through Bayes formula:

$$P[T|G,E] = \frac{P[G|T,E]P[T|E]}{P[G|E]},$$

they would be equivalent only under the knowledge of distributions $P[T|E]$ and $P[G|E]$, which are not available. In this section, we illustrate why the survival analysis models developed for genome-wide nested case-control design, in particular the Cox model, cannot be directly adapted to the post-genome design, and we present an alternative approach based on a gene-by-gene model.

## 3.1 Are survival analysis models relevant for post-genome design?

**Semi-parametric models.** Survival analysis models parameterize a time of failure $T$ (in our context, the follow-up time) as a function of a vector of variables $X$. The analysis in nested case-control design has been mainly focused on semiparametric models. Zheng et al [15] propose a procedure of estimation adapted to the nested case-control design for a very flexible model including additive and multiplicative proportional effects. The procedure for the additive part of the model requires knowledge of the history of the covariate, i.e. the value of the covariates at any time before $T$. This information is not available for transcriptomics variables, which exclude the use of the additive hazard model in our context. Conversely, multiplicative hazard models can be estimated from a unique observation of the covariates. Let us examine their relevance for post-genome data.

**Cox proportional hazard model for nested case-control design.** The Cox model (see Figure 4) is one of the most popular model in prospective nested-case control studies. The hazard rate, defined as the risk of occurrence of an event just after a time $t$ given that it did not occurs before $t$ is the product of a baseline hazard rate function and the exponential of a linear combination of covariates. This model, originally developed by Cox [2] does not rely on a biological modeling, but has been designed to address a specific statistical goal: the relative risks can be computed from right-censored observations without estimating the baseline, via a partial likelihood based on the ranking of the failure times.

Thomas [5] adapted the partial likelihood technique to nested case-control design, and theoretical properties of his estimator were later formally established using counting processes and martingale theory [16]. In this context, the sampled risk set at the failure time of a case $i$ consists of the matched controls and the follow-up time is not involved in the partial likelihood. Therefore, Cox model estimation is equivalent to a simple logistic regression between cases and controls and the coefficient associated with a covariate $X^{(j)}$ measures the ability of the covariate to discriminate between cases and controls (see [17] for details).

**Cox proportional hazard model for post-genome design.** Consider the Cox model applied to the post-genome cohort data $(T_i, \Delta G_i)$ with $T_i$ the follow-up time and $\Delta G_i$ the difference of log-expression for the case-control pair $i$. The coefficients $\beta_g$ affecting the partial likelihood for a nested case-control design correspond to genes which are differentially expressed between cases and controls. This may includes genes which are constantly differentially expressed as well as genes whose expression changes before diagnosis. Therefore, the Cox model does not appear efficient to discriminate between these two behaviours.

**Cox model with time-varying coefficients.** In the past twenty years, Cox model has generated a huge number of developments, which have provided important generalizations from the original model. In particular, the introduction of time-varying coefficients offers a more flexible framework [6] but its implementation in nested-case control studies remains limited. Liu et al [18] propose an estimator based on a kernel-weigthed local polynomial approach. Intuitively, the estimation of $\beta(t)$ is based on the pairs with follow-up times in the close vicinity of $t$. The non-constant coefficients corresponds to genes with a non-multiplicative effect and do not specifically target genes whose expression changes over time. Therefore, the Cox model with time varying coefficients presents the same pitfalls as the classical Cox model: it does not provide a discrimination between the genes associated with cancer and the genes whose expression changes before diagnosis.

**Penalization procedures.** In genome-wide studies, a penalization procedure is necessary to handle the large dimension covariates [3, 4]. It extracts a restricted set of genes which have the highest effect on the partial log-likelihood. Thus, on post-genome data, penalization would provide a restricted set of genes which offers the best discrimination between cases and controls. Therefore, genes whose expression changes before diagnosis could be eliminated if they display a poorer discrimination ability than genes constantly differentially expressed.

**Summary.** We have examined the relevance of classical semi-parametric models for the post-genome design. Additive hazard rate models are excluded in a context where a single measurement of gene expression is available. Multiplicative hazard rate models in a nested case-control design estimate the ability of a gene to discriminate between cases and controls, and do not appear efficient to specifically target genes whose expression changes over time.

## 3.2 An alternative approach: parametrization of gene trajectories

We propose a framework to directly handle the changes in gene expression during the carcinogenic process, if any, based on transcriptomics data from the post-genome cohort. Contrary to the models developed for genomics data in GWAS, transcriptomics are not considered as risk factors but as markers of the carcinogenic process and exposures.

The expression of each gene is parameterized as a function of time to diagnosis and exposures. For each gene $g$ and each case $i$:

$$G_{i,g}^{\text{case}} = f(T_i, E_i | \Theta_g) + \varepsilon_{i,g} \qquad (3)$$

for some parametric function $f(.|\Theta_g)$. The exposures vector $E_i$ includes lifestyle or environmental variables which affect gene expression, regardless of their association with cancer. Contrary to survival analysis models, the error term $\varepsilon_{i,g}$ does not model the random occurrence of an event (diagnosis of cancer), but accounts for individual variations of gene expression. As described in Section 2.2, the controls are used as a reference to correct age and technical effects. Thus, we consider the following model for the case-control differential gene expression $\Delta G$:

$$\Delta G_{i,g} = h(T_i, \Delta E_i | \Theta_g) + \varepsilon'_{i,g} \qquad (4)$$

for some parametric function h(.|Θ$_g$). The "difference of exposure" $\Delta E_i$ for the case-control pair *i* can be equal to the difference for a continuous exposure, or to ($E_i^{case}$, $E_i^{control}$) for a categorical exposure. For each gene *g*, a p-value for the association between time to diagnosis and gene expression can be computed from model (4). Finally, a multiple testing procedure is implemented to control the type I error [19, 20].

Model (4) represents a flexible framework which can include various biological assumptions. The following examples illustrate the potentialities of a gene-by-gene parameterization to explore functional changes in gene expression. The corresponding statistical models are provided in the Appendix.

**(a) Additive exposure effects and parametric time effect.** The differential expression of each gene is modelled as the sum of a linear effect of a vector of exposures and a carcinogenesis effect parameterized as a function of time. In particular the time effect could be modelled by a "hockey stick" function: the gene starts linearly over- or under-expressing at a certain time before diagnosis. Contrary to the Cox model, this model make the distinction between a constant case-control difference and changes in gene expression before diagnosis.

**(b) Non-parametric time effect.** The unknown shape of the gene trajectory can be handled with a non-parametric estimator of the time effect. In particular, isotonic regression [21] allows testing of a monotonic effect of time on gene expression.

**(c) Exposure-driven functional model of carcinogenesis.** The gene-by-gene parametrization allows inclusion of more advanced hypotheses on gene-environment interactions. In particular, the assumption of carcinogenesis mechanisms driven by exposures [22] could be tested via a parameterization including interaction terms between time and carcinogen exposures.

The implementation of gene-by-gene trajectory models requires choices based on epidemiological and biological knowledge. First of all, a set of relevant exposures that affect gene expression has to be determined. A cross-sectional study from an independent set of controls could provide a restricted set of exposures which present the strongest association with gene expression. Alternatively, meta-variables expressed as linear combinations of a large vector of exposures could be extracted by a principal components analysis. Moreover, regarding breast cancer, a stratification of the cases based on cancer stage or receptor status would be necessary to account for heterogeneity of the carcinogenic process. A modeling including mechanisms common to all cases, as well as specific mechanisms for a given subgroup of cases could be implemented.

# 4 Conclusion

Exploration of the carcinogenic process on transcriptional outputs at the epidemiological level opens up new horizons in functional analysis and causal inference, and requires a new design together with adequate analysis procedures. Despite similarities with the more classical prospective GWAS, moving from genomics to transcriptomics addresses different goals and the relevance of canonical statistical methods for GWAS in these novel studies is questioned. Whereas prospective GWAS explore the effects of genomics and environmental variables on the time of occurrence of cancer via survival analysis models $P[T|G,E]$, post-genome studies explore changes in the transcriptome during the carcinogenic process conditional on exposures: $P[G|T,E]$.

The post-genome cohort includes transcriptomics measurements together with environmental exposures in a nested case-control prospective design. The transcriptional outputs observed in different individuals at different times before diagnosis, the trajectories, could be adjusted for the effects of exposures as a source of variation between individuals.

The survival analysis models classically used in prospective GWAS, in particular the Cox model, have proven to be highly efficient for risk estimation and survival prediction. Nevertheless, we suggest that they are not appropriate to explore the functional changes in gene expression during the carcinogenic process. Indeed, in a genome-wide context, these models does not specifically target genes whose expression changes before diagnosis, but genes which offer the best discrimination between cases and controls. Therefore we propose to directly model the trajectory of the expression of each gene in a model including time to diagnosis and exposures. This flexible approach allows to model and test various relationships between gene expression and exposures in the carcinogenic process, and offers a direct interpretation in terms of carcinogenesis mechanisms.

## Acknoledgements

Grant: ERC-2008-AdG 232997-TICE "Transcriptomics in cancer epidemiology".

# Figures

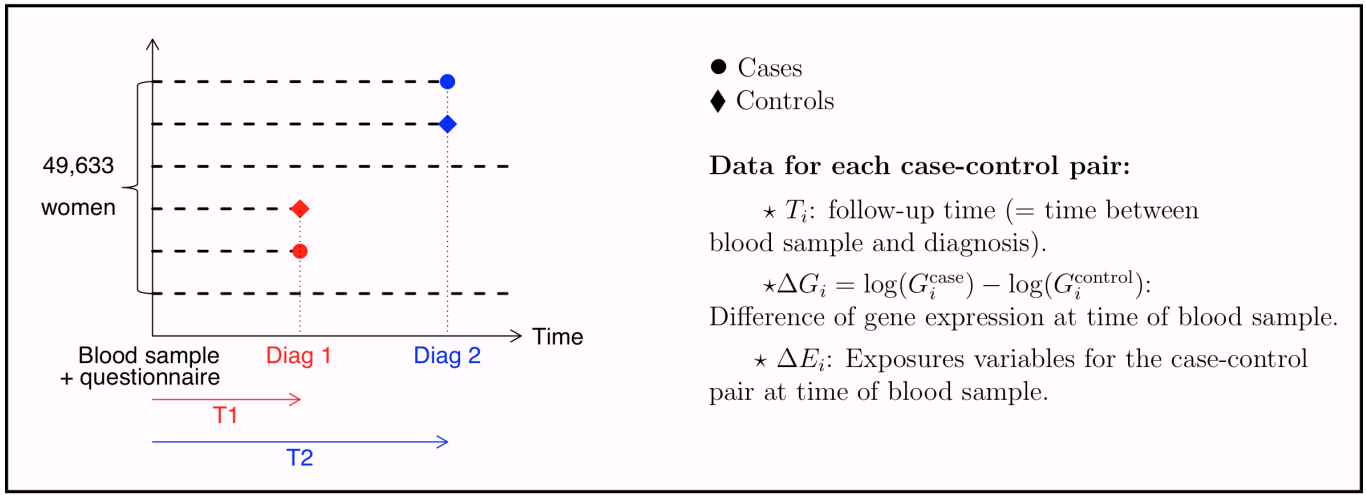

Figure 1: Design of the NOWAC post-genome cohort

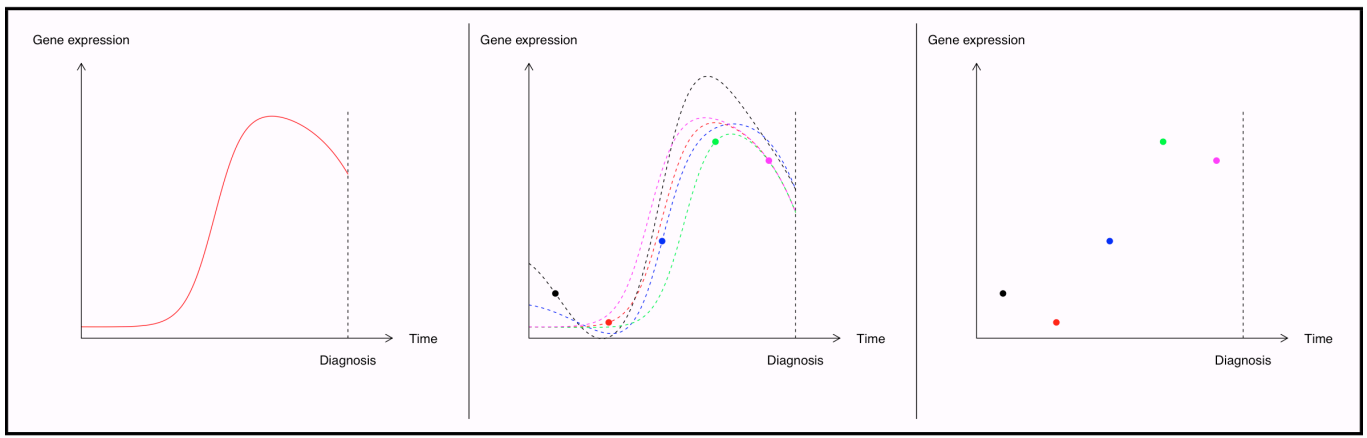

Figure 2: **Gene trajectory: hypothetical example.** The first column presents the trajectory of a gene *g* for one individual. The second column displays the trajectories of *g* for five individuals (dotted lines), as well as measurements of gene expression at different times before diagnosis (dots). The measurements of gene expression isolated on the third column offer a good recovery of the trend in the trajectories.

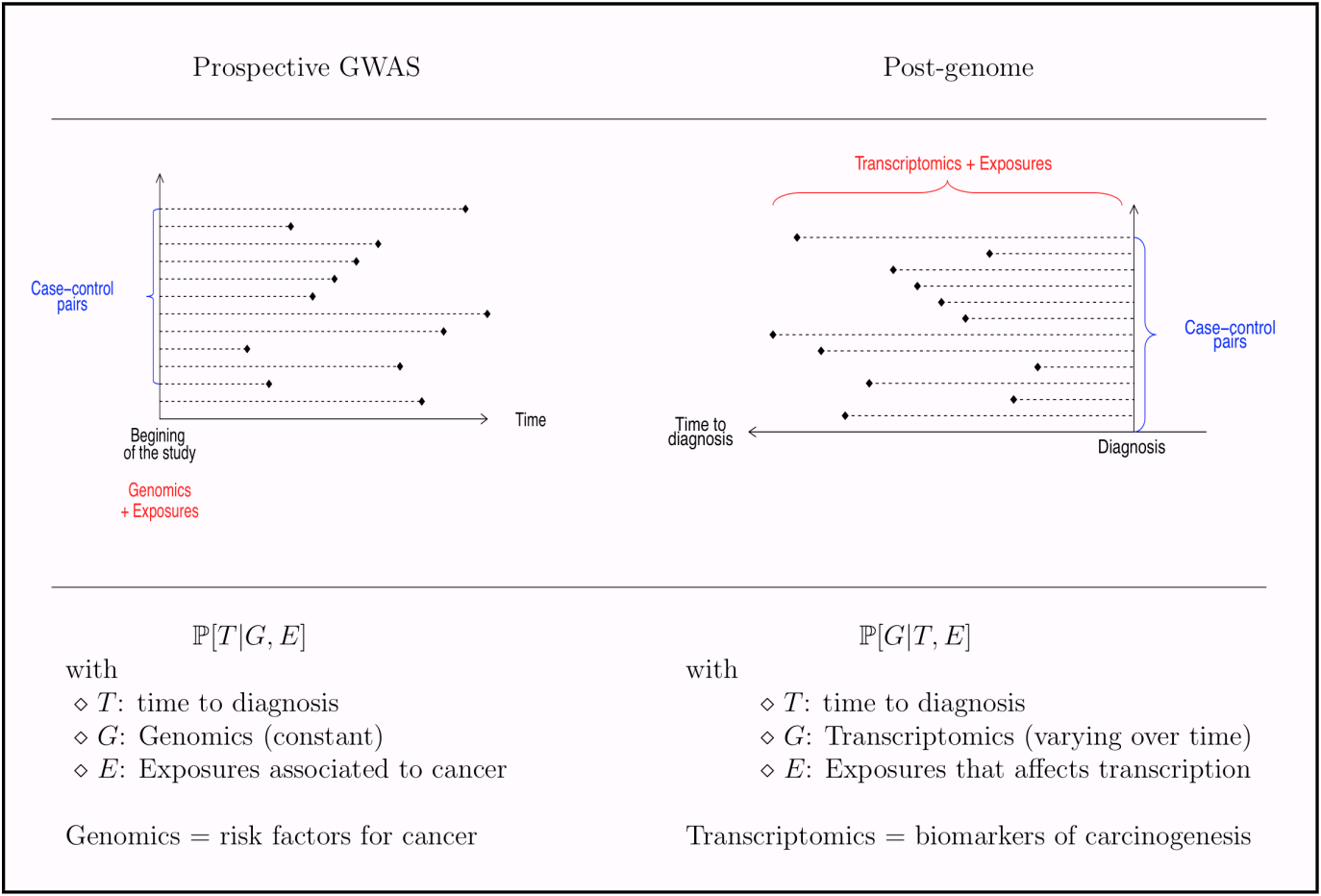

Figure 3: From prospective GWAS to post-genome design.

## Cox model

- Cox proportional hazard rate: for vector of covariates $X$

$$\lambda(t|X) = \lambda(t) \exp\left(\langle \beta_0, X \rangle\right)$$

with $\langle .,. \rangle$ the canonical scalar product.

- Partial likelihood for nested case-control design:

$$L(\beta) = \prod_{i \text{ case-control pair}} \left(1 + \exp\left(\langle \beta, \Delta X_i \rangle\right)\right)^{-1}.$$

with $(\Delta X_1, \ldots, \Delta X_n)$ the differences of covariates between the matched case-control pairs

- Cox model with time-varying coefficients:

$$\lambda(t|X) = \lambda_0(t) \exp\left(\langle \beta(t), X \rangle\right).$$

- Penalized log-likelihood procedure:

$$\widehat{\beta} = \arg\min\left(-\log L(\beta) + \text{pen}(\beta)\right)$$

for some penalization function pen(.).

Figure 4

# Supplementary material

We present the statistical models corresponding to the examples of gene-by-gene models described in Section 3.2.

**(a) Additive exposure effects and parametric time effect.**

$$\Delta G_{i,g} = \alpha_0^g + \langle \alpha_1^g, \Delta E_i^{(1)} \rangle + \alpha_2^g T_i \mathbb{I}(T_i > t^g) + \varepsilon'_{i,g}.$$

The coefficient $\alpha_0^g$ accounts for a constant difference between cases and controls along the follow-up time of the study, which can indicate changes in gene expression occurring earlier in the carcinogenic process, or exposure effects which have not been included in vector $E^{(1)}$. The coefficient vector $\alpha_1^g$ corresponds to the effects of exposures $E^{(1)}$ on the expression of gene $g$. Finally, $\alpha_2^g$ accounts for time effect: a significantly non-zero coefficient $\alpha_2^g$ assesses an effect of carcinogenesis on gene $g$.

**(b) Non-parametric time effect.** The gene trajectory is estimated through a non-parametric estimator $\varphi$ of the time effect:

$$\Delta G_{i,g} = \alpha_0^g + \langle \alpha_1^g, \Delta E_i^{(1)} \rangle + \varphi(T_i) + \varepsilon'_{i,g}$$

**(c) Exposure-driven functional model of carcinogenesis.** A model including interactions between the follow-up time and an exposure $E^{(2)}$ accounts for the involvement of a gene $g$ in the carcinogenic process under the exposure $E^{(2)}$:

$$\Delta G_{i,g} = \alpha_0^g + \langle \alpha_1^g, \Delta E_i^{(1)} \rangle + \varphi(T_i) + E_i^{(2),\text{case}} \psi(T_i) + \varepsilon'_{i,g}$$

Function $\varphi$ stands for a global effect of carcinogenesis on all cases, whereas $\psi$ accounts for a driven by exposure effect, affecting the gene expression of pairs whose case is under exposure $E^{(2)}$.